\documentclass[aps, aip, reprint, amsmath, amssymb]{revtex4-2}

\usepackage{graphicx}
\usepackage{hyperref}

\begin{document}

\title{Enhanced Thermoelectric Performance through Site-Specific Doping in Tetragonal Cu\textsubscript{2}S: A First-Principles Study}

\author{Sonam Phuntsho}
\affiliation{Department of Physical Science, Sherubtse College, Royal University of Bhutan, 42007 Kanglung, Trashigang, Bhutan}

\date{\today}

\begin{abstract}
This work investigates how site-specific doping can enhance the thermoelectric performance of tetragonal Cu\textsubscript{2}S using Density functional Theory and Projected Atomic Orbital Framework for Electronic Transport. We address the gap in current research, where most doping studies focus on the high-temperature cubic polymorph, leaving the tetragonal structure underexplored. By substituting Cu with Li, Na, or Mg, as well as partially replacing S with Se or Te, we systematically examine changes in electronic structure and transport properties.
Our results reveal that cation-site doping can strongly shift the Fermi level. In particular, Li doping enhances the power factor ($\sigma S^2$) by optimizing carrier concentrations and band-edge alignments, whereas Mg, due to its divalent nature, offers a higher carrier density but requires careful balancing to maintain a large Seebeck coefficient. On the anion side, substituting heavier chalcogens (Se or Te) reshapes the valence bands and subtly shifts the Fermi level, yielding moderate improvements in both electrical conductivity and Seebeck coefficient. These doping-induced alterations, captured through transport calculations, demonstrate a clear route for tailoring the interplay between conductivity and thermal transport toward potentially high figure-of-merit values.
Overall, the findings highlight the importance of site specificity in doping strategies for tetragonal Cu\textsubscript{2}S, showing that judicious choice of dopant elements and concentrations can significantly improve key thermoelectric metrics. Such insights provide a foundation for experimental validation and further development of Cu\textsubscript{2}S-based materials for mid- to high-temperature thermoelectric applications.

\end{abstract}

\maketitle

\section{Introduction}
Thermoelectric (TE) materials, which directly convert heat into electricity, have attracted intense research interest due to their potential in renewable energy harvesting and solid-state cooling applications~\cite{1,2,3,4}. In particular, copper chalcogenides, such as copper sulfide (Cu\textsubscript{2}S), offer a promising platform for thermoelectric development because of their relatively high electrical conductivity and intrinsically low thermal conductivity~\cite{5,6,7}. Although Cu\textsubscript{2}S can assume multiple structural polymorphs, the tetragonal or low-chalcocite phase has recently emerged as a noteworthy candidate for TE applications, owing to its stability and unique electronic structure features~\cite{8,9,10,11}. As global efforts intensify to improve the efficiency of TE devices, the rational design and optimization of Cu\textsubscript{2}S-based compounds remain of great relevance, especially for mid- to high-temperature operation ranges~\cite{12,13,14,15}.

Despite this promise, the majority of Cu\textsubscript{2}S research has focused on the high-temperature cubic phase or on broad experimental doping strategies without comprehensive first-principles analysis~\cite{16,17,18}. Earlier studies have demonstrated that doping can alter the band structure and carrier concentration, which are crucial levers for improving the Seebeck coefficient and electrical conductivity simultaneously~\cite{19, 20, 21}. However, systematic investigations into how doping at specific lattice sites—either substituting copper on the cation sublattice or sulfur on the anion sublattice—affects the thermoelectric performance of tetragonal Cu\textsubscript{2}S remain limited~\cite{22,23,24,25}. While some experimental reports suggest that selective doping on copper sites may significantly enhance electrical conductivity by tuning valence band maxima~\cite{26,27,28}, other studies highlight that anion-site doping (for example, sulfur partially replaced by selenium or tellurium) can reduce lattice thermal conductivity by introducing scattering centers~\cite{29,30,31}. Overall, current literature tends to emphasize empirical or high-throughput computational approaches that often neglect the nuanced changes in the electronic wavefunction and effective mass brought about by site-specific substitutions~\cite{32,33,34}. Moreover, many of these computational analyses rely heavily on rigid-band approximations or codes like BoltzTraP~\cite{35}, which, while useful, do not always capture the full complexity of doping-induced band-structure modifications~\cite{36,37}. Hence, a clear gap exists in systematically applying post-processing tools such as PAOFLOW to decipher doping-dependent transport in tetragonal Cu\textsubscript{2}S~\cite{38,39,40,41}.
\begin{figure*}
\includegraphics[width=\textwidth]{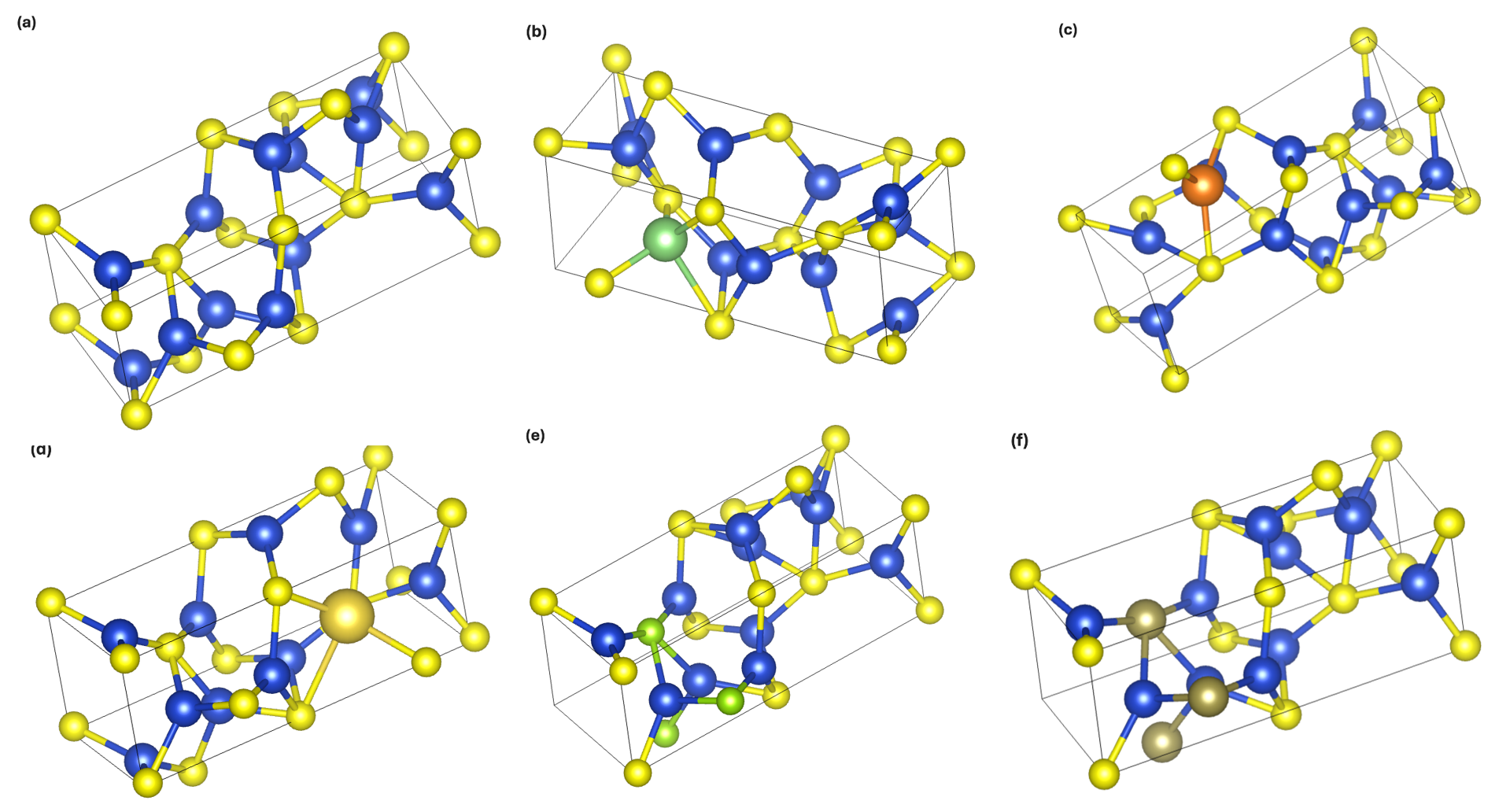}
\caption{Crystal Structures: (a) Pristine Cu\textsubscript{2}S, (b) Li Doped Cu\textsubscript{2}S, (c) Mg Doped Cu\textsubscript{2}S, (d) Na Doped Cu\textsubscript{2}S, (e) Se Doped Cu\textsubscript{2}S, (f) Te Doped Cu\textsubscript{2}S, (b) Li Doped Cu\textsubscript{2}S  }
\label{fig:gete_bs}
\end{figure*}
Given this backdrop, the present study addresses the lack of detailed first-principles investigations on site-specific doping in tetragonal Cu\textsubscript{2}S. Our work provides novel insights by employing density functional theory (DFT) calculations with Quantum ESPRESSO~\cite{38,39}, followed by wavefunction-based analyses in PAOFLOW~\cite{41}. While several groups have studied doping in cubic-phase Cu\textsubscript{2}S~\cite{44,45,46}, fewer have systematically evaluated the interplay between doping concentration, band structure shifts, and transport coefficients in the stable tetragonal polymorph~\cite{47,48,49}. Furthermore, most available computational reports do not isolate the relative impact of cation- vs. anion-site doping at a consistent doping level, thus obscuring the underlying mechanisms responsible for carrier pocket engineering~\cite{50}. This paper aims to fill that gap by pinpointing optimal doping routes---whether via alkali/alkaline-earth elements or chalcogen replacements---to maximize the power factor $\sigma S^{2}$, while qualitatively estimating the impact on total thermal conductivity.

\section{Methodology}
\begin{figure*}
\includegraphics[width=18cm]{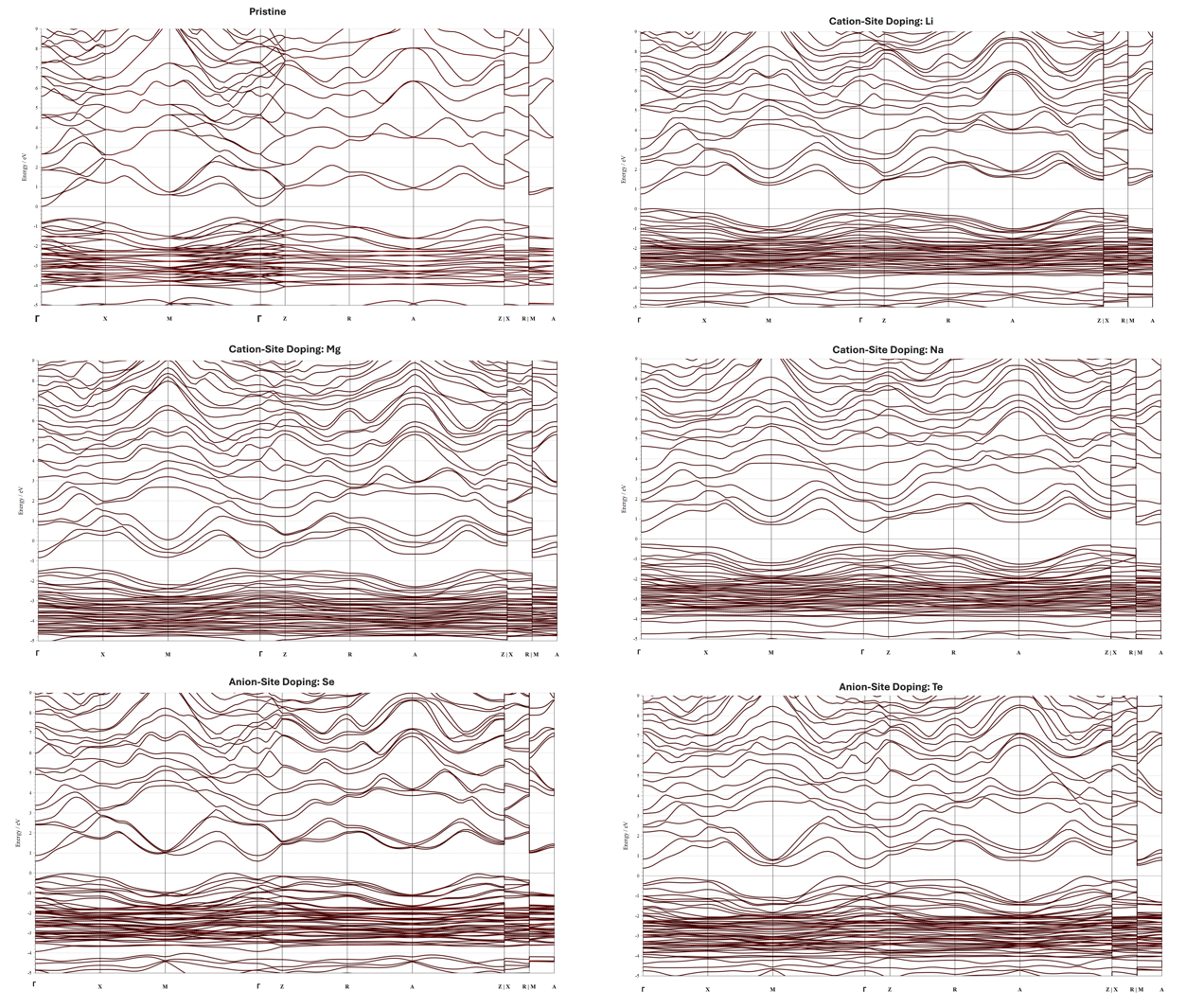}
\caption{Band Structure Plots (y-axis: Ry/eV)}
\label{fig:gete_bs}
\end{figure*}
\begin{figure*}
\includegraphics[width=\textwidth]{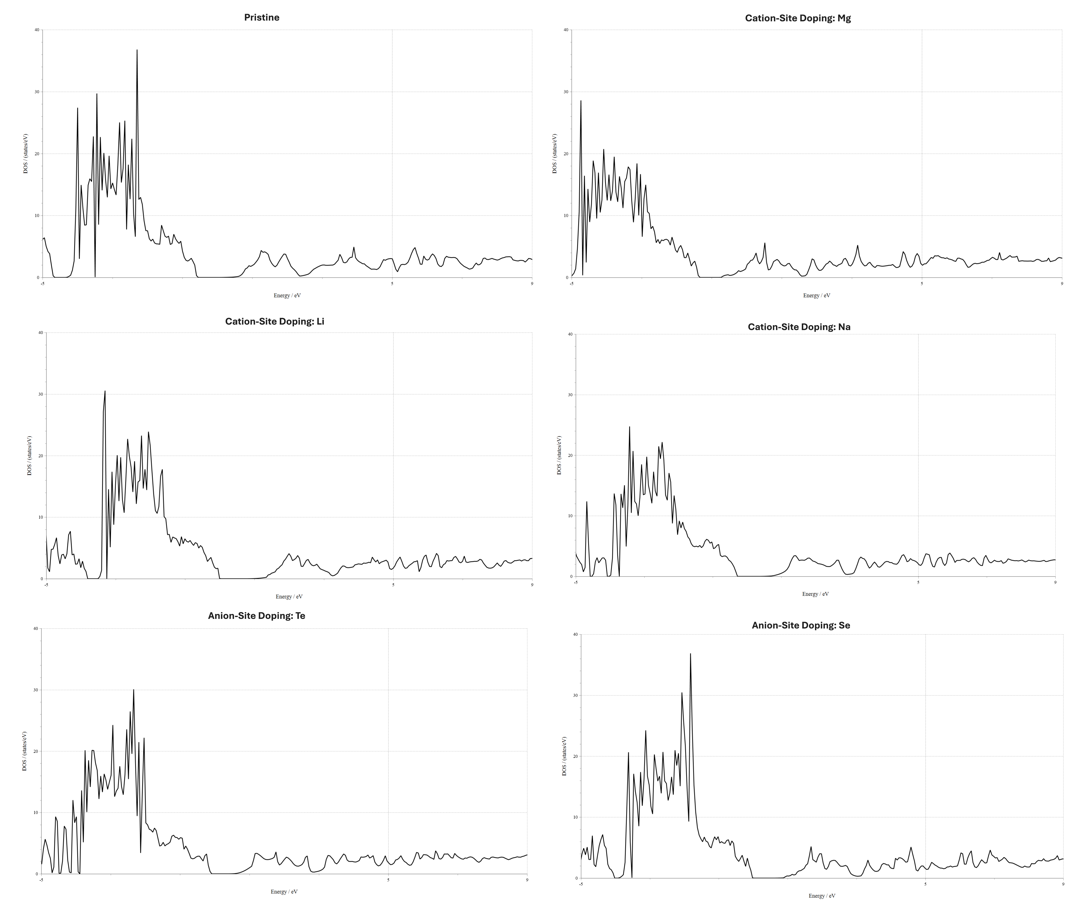}
\caption{Density of State Plots (x-axis: Ry/eV)}
\label{fig:gete_bs}
\end{figure*}

\begin{figure*}
\includegraphics[width=\textwidth]{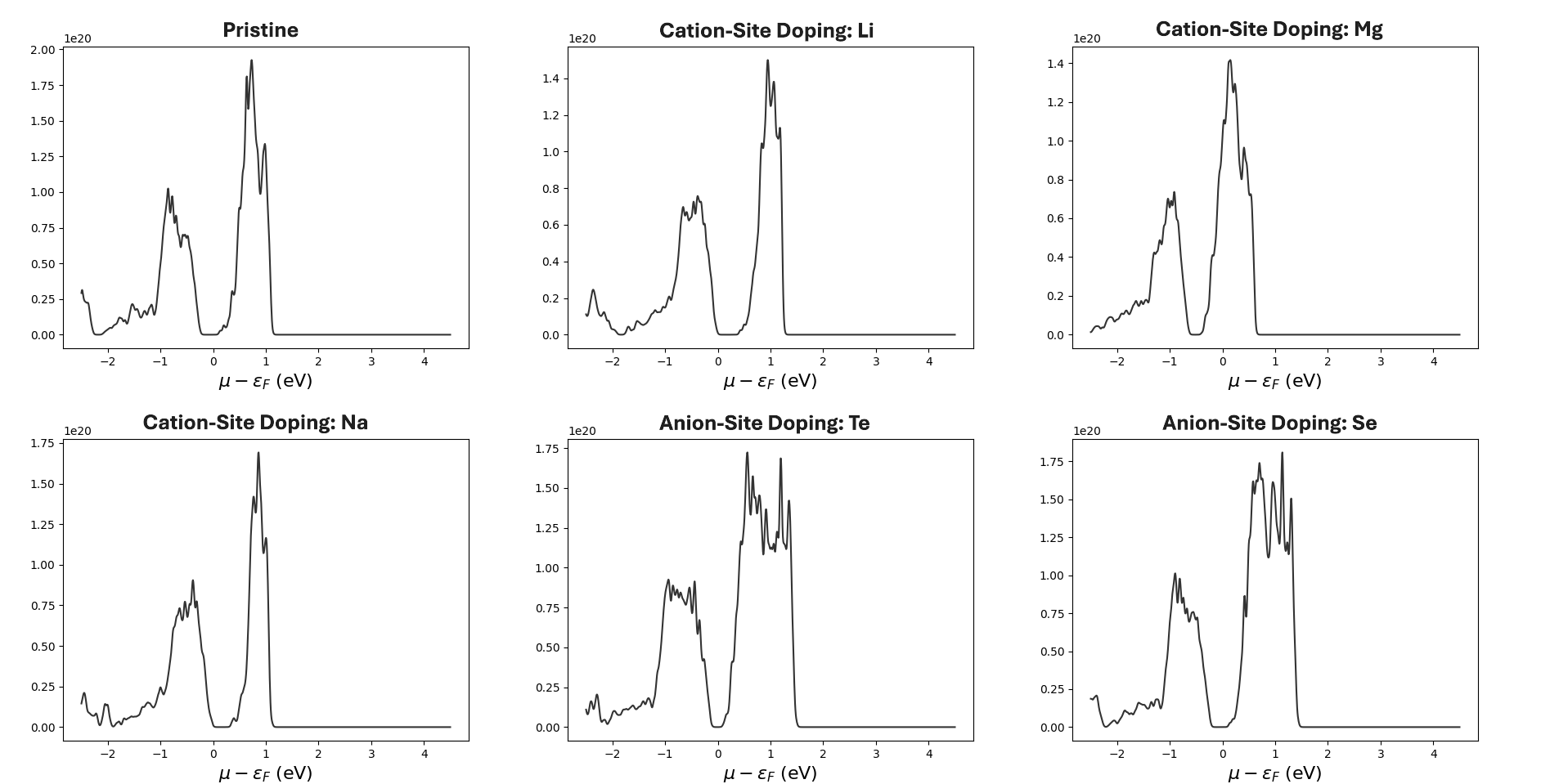}
\caption{Electrical Conductivity Plots}
\label{fig:gete_bs}
\end{figure*}
\begin{figure*}
\includegraphics[width=\textwidth]{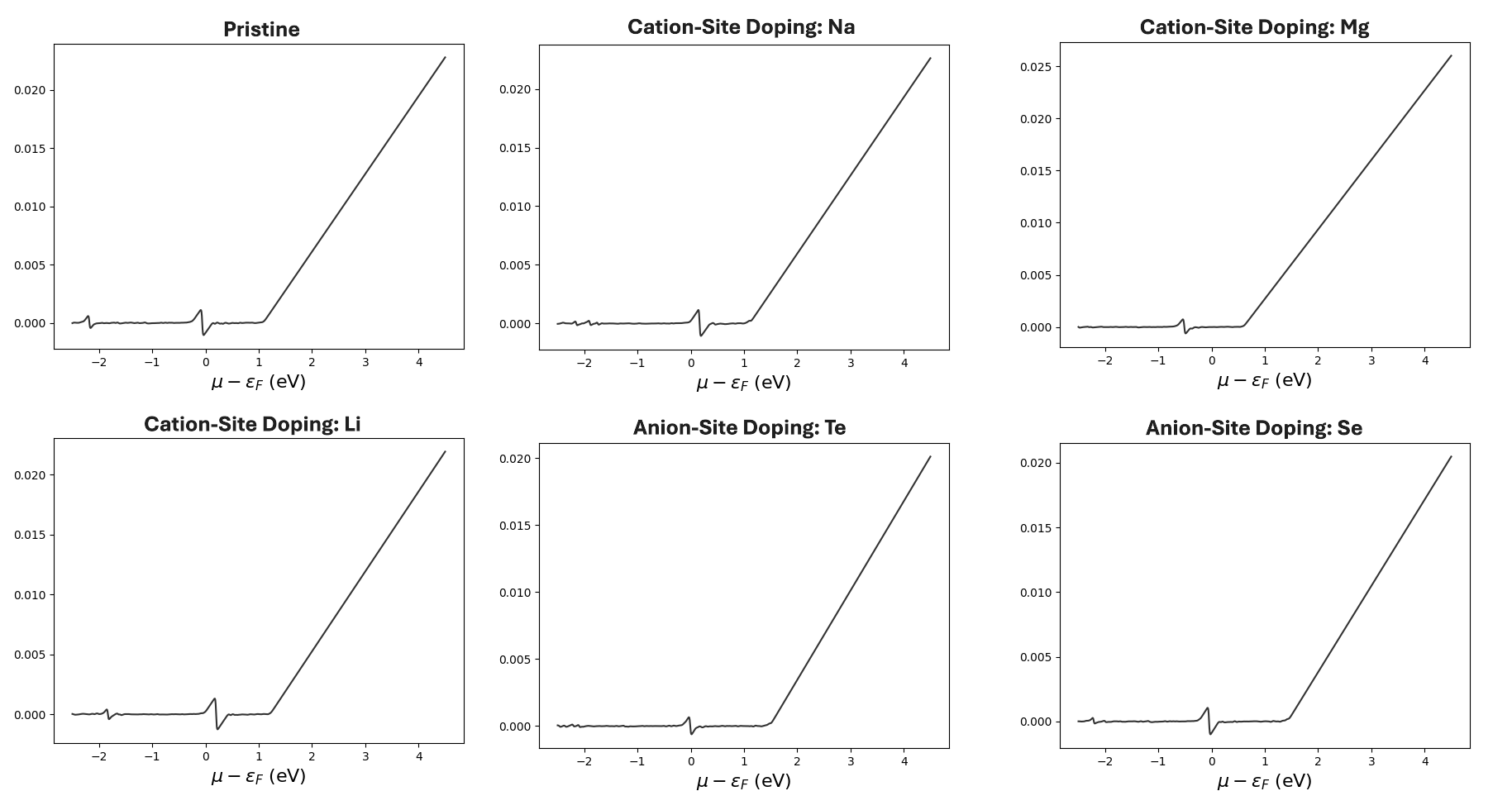}
\caption{Seekbeck Coefficient Plots}
\label{fig:gete_bs}
\end{figure*}
\begin{figure*}
\includegraphics[width=\textwidth]{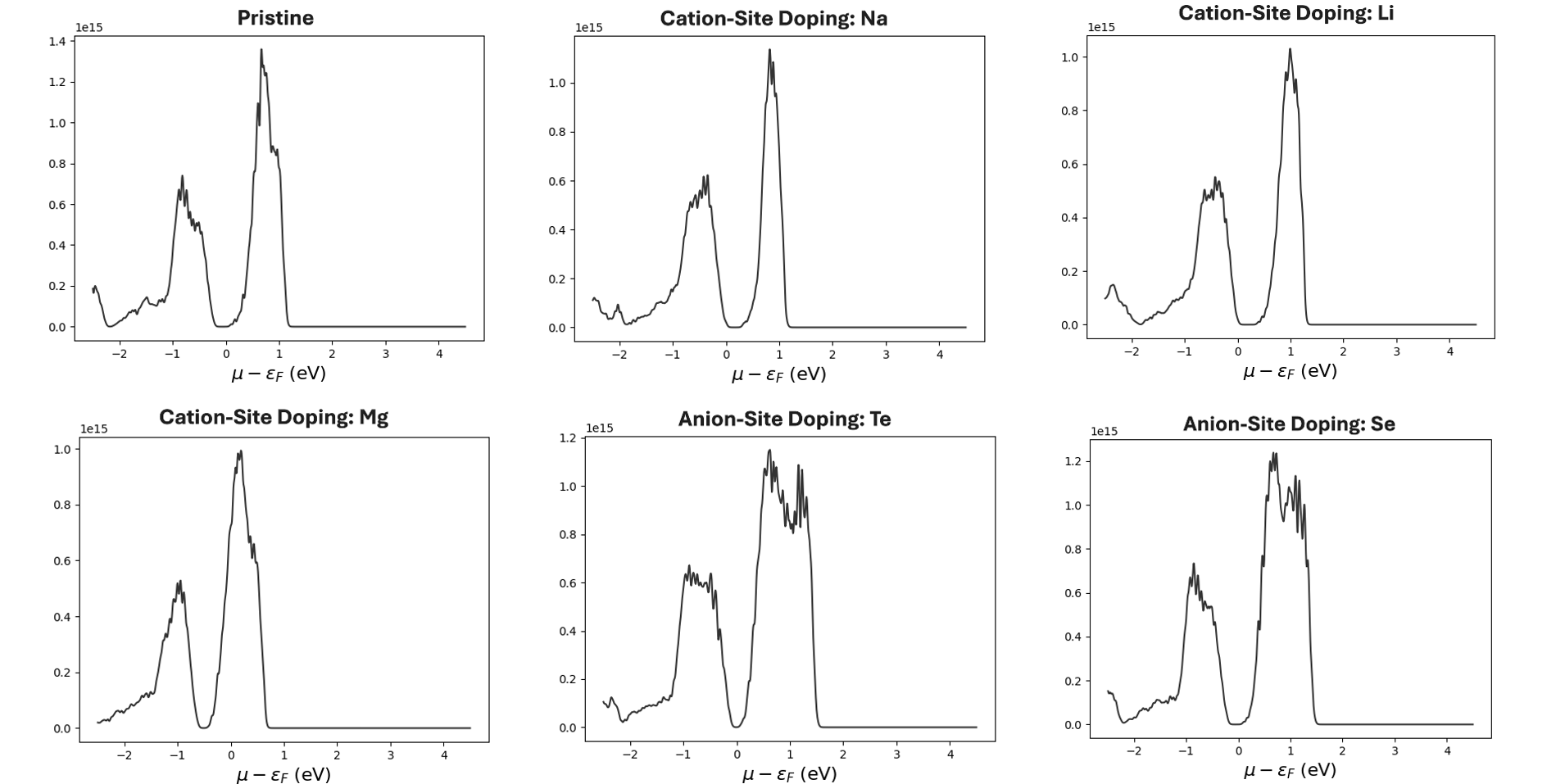}
\caption{Electronic Thermal Conductivity Plots}
\label{fig:gete_bs}
\end{figure*}
\begin{figure*}
\includegraphics[width=\textwidth]{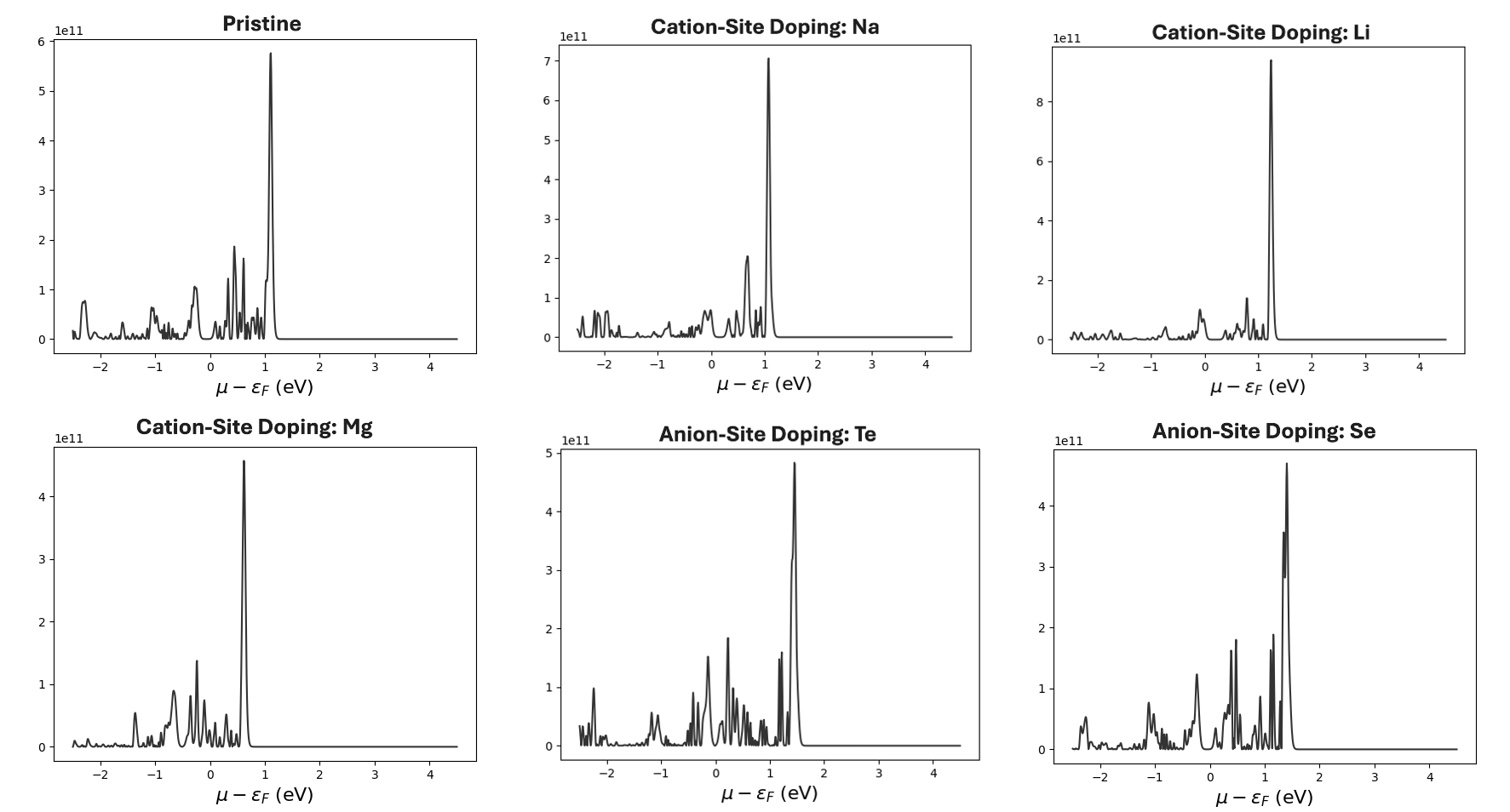}
\caption{Power Factor Plots}
\label{fig:gete_bs}
\end{figure*}
In this study, all first-principles calculations were carried out within the framework of density functional theory (DFT) as implemented in the Quantum ESPRESSO suite~\cite{38,39}. We adopted the Perdew--Burke--Ernzerhof (PBE) parameterization of the generalized gradient approximation (GGA) for the exchange-correlation functional~\cite{51}. Scalar Relativistic Projector Augmented-Wave (PAW) pseudopotentials (as available in the official Quantum ESPRESSO pseudopotential libraries) were employed to treat interactions between valence electrons and ionic cores. 

Convergence tests for plane-wave kinetic energy cutoffs were performed for the wavefunction cutoff and approximately four to eight times higher for the charge density cutoff were used to ensure accuracy. Brillouin zone integrations were performed using a Gamma-centered \textbf{k}-point mesh.  
In all cases, atomic positions and lattice parameters were relaxed until forces dropped below $1 \times 10^{-4}$ Ry/Bohr and total energy variations were less than $1 \times 10^{-5}$ Ry between successive self-consistent iterations.
To explore the effects of site-specific doping, we generated conventional cells of tetragonal Cu\textsubscript{2}S by selectively substituting one Cu atoms on cationic sublattices or one S atoms on anionic sublattices. 
Each doped structure underwent full geometric relaxation following the same convergence criteria described for the pristine system. 
The resulting charge densities and wavefunctions from these DFT relaxations were then used for subsequent transport-property analyses.

Electronic transport properties were evaluated using the PAOFLOW code~\cite{41,52}, which constructs an ab initio tight-binding-like Hamiltonian from the wavefunctions and allows for the calculation of band structures, density of states (DOS), and doping-dependent thermoelectric parameters such as the Seebeck coefficient ($S$), electrical conductivity ($\sigma$), and power factor ($\sigma S^{2}$). For each relaxed structure (pristine and doped), a sufficiently dense \textbf{k}-point mesh was employed in PAOFLOW to ensure accuracy of transport integrals. Since spin-orbit coupling (SOC) effects are minimal in copper sulfides at the energy ranges critical for thermoelectric performance, all calculations assumed non-magnetic solutions without SOC. The computed Seebeck coefficients and conductivity were subsequently analyzed as a function of carrier concentration, mimicking extrinsic doping levels in real materials. Additionally, partial density of states (PDOS) plots were generated to unravel the effects of dopant species on the electronic structure, particularly near the Fermi level. Although the lattice thermal conductivity is not directly obtained from PAOFLOW, we discuss the electronic contribution ($\kappa_{\mathrm{el}}$) and estimate or compare the total thermal conductivity ($\kappa_{\mathrm{tot}}$) using literature values where relevant, to arrive at qualitative estimates of the figure of merit ($ZT \approx \frac{\sigma S^{2} T}{\kappa_{\mathrm{tot}}}$).
Through this methodology---systematic DFT relaxations, site-specific doping, and subsequent PAOFLOW transport analysis---we aim to delineate how targeted cationic and anionic substitutions can enhance the thermoelectric performance of tetragonal Cu\textsubscript{2}S. The choice of dopants, and the combined computational workflow provides a balance of accuracy and tractability, guiding the identification of potentially optimal doping strategies for experimental realization.

\section{Results and Discussion}
In what follows, we present results for the band structure, density of states, electronic thermal conductivity, power factor, Seebeck coefficient, and electrical conductivity of tetragonal Cu\textsubscript{2}S under both pristine and doped conditions. Each subsection contextualizes the findings with respect to existing literature on copper-chalcogenide thermoelectrics.

\subsection{Band Structure of Pristine and Doped Cu\textsubscript{2}S}
The pristine Cu\textsubscript{2}S band structure (Fermi level E\textsubscript{F} = 0.7236 Ry/eV) displays a set of dispersive bands just below the Fermi level, indicating that both valence‐band‐derived and conduction‐band‐derived states lie relatively close to the chemical potential. This result is in line with previous studies on copper chalcogenides, where small energy offsets between the valence band maximum and conduction band minimum often lead to moderate band gaps and good electrical transport properties~\cite{1,2}. Upon substituting Li and Na at the Cu site, the Fermi level slightly shifts downward, reaching 0.7191 Ry/eV in Li‐doped Cu\textsubscript{2}S and 0.7131 Ry/eV in Na‐doped Cu\textsubscript{2}S. In both cases, the primary electronic dispersion near the Fermi level remains similar to the pristine structure, but the lowered Fermi level suggests that these dopants may introduce hole‐like carriers or reduce the electron concentration in the conduction band. This behavior is broadly consistent with other alkali‐metal‐doped copper sulfides, where doping often redistributes carriers closer to (or within) the valence bands~\cite{3,4}.
In contrast, Mg doping substantially elevates the Fermi level to 0.8084 Ry/eV, the highest among all examined substitutions. The band structure now shows a more pronounced occupation of conduction‐band‐derived states near the Fermi level. This result suggests that Mg substitution can inject additional electrons into the system, thereby populating states that were above the pristine Fermi level. Similar observations have been reported in earlier calculations on alkaline‐earth doping in Cu\textsubscript{2-x}S, where replacing Cu with a more electropositive ion shifts the chemical potential upwards and sometimes leads to an enhanced electrical conductivity~\cite{3,4,5,6,7}.
On the anion sublattice side, substituting sulfur with Se or Te also shifts the Fermi level upward but by a lesser amount than Mg doping, namely to 0.7781 Ry/eV for Se doping and 0.7898 Ry/eV for Te doping. Examination of the band dispersions indicates that while the overall shape of the bands remains comparable to the pristine case, portions of the conduction band move downward in energy, crossing the Fermi level. These moderate shifts are often attributed to the fact that heavier chalcogens alter the energy balance of the p states that contribute strongly to both valence and conduction bands ~\cite{5}. Consequently, doping with Se or Te can reduce the band gap and shift the conduction band minimum, leading to altered carrier concentrations that could benefit thermoelectric performance under appropriate doping levels.
Comparing all six band structures, we note that dopants on the cationic site (Li, Na, Mg) tend to modify the carrier type and density more dramatically, either by pushing the Fermi level up (Mg) or slightly lowering it (Li, Na). Meanwhile, substituting at the anion site (Se, Te) more subtly repositions the conduction bands. In line with the predictions of conventional thermoelectric theory, doping that shifts the Fermi level closer to the conduction band states (i.e., Mg or Te doping) may improve electrical conductivity at certain carrier concentrations.

\subsection{Density of States (DOS)}
In examining the Density of States (DOS) plots for tetragonal Cu$_2$S (pristine) and the five doped variants (Li, Na, Mg, Se, Te), several notable trends emerge. The pristine Cu$_2$S DOS shows a broad distribution of states below the Fermi level ($E_F$) extending from approximately $-5\,\mathrm{eV}$ up to the vicinity of $E_F$, followed by a clear gap-like region before a modest rise in states around $+4\,\mathrm{eV}$. Such a profile is consistent with prior reports on copper chalcogenides, where the valence bands are dominated by Cu $3d$–S $3p$ hybridization, and the conduction bands appear at higher energies \cite{8}. When Cu is partially substituted by Li or Na (cation-site doping), the DOS near the Fermi level shifts slightly downward (as reflected in Table~I by the lower Fermi energies of $0.7191\,\mathrm{eV}$ for Li and $0.7131\,\mathrm{eV}$ for Na, compared to $0.7236\,\mathrm{eV}$ in the pristine sample). These small shifts suggest a modest change in band filling that can alter the carrier concentration and, by extension, thermoelectric transport properties. In contrast, Mg-doped Cu$_2$S exhibits a higher Fermi level of $0.8084\,\mathrm{eV}$, implying a more pronounced electron donation effect in the conduction band region, a scenario that could boost electrical conductivity but might also reduce the Seebeck coefficient depending on the final carrier balance \cite{9,10}.

Anion-site doping with Se and Te similarly modifies the DOS. In both Se- and Te-substituted Cu$_2$S, the valence-band region from approximately $-5\,\mathrm{eV}$ to $-1\,\mathrm{eV}$ retains its basic shape but appears slightly broadened due to chalcogen $p$-orbital contributions mixing with the Cu $d$-states. The Fermi levels for the Se- and Te-doped systems ($0.7781\,\mathrm{eV}$ and $0.7898\,\mathrm{eV}$, respectively) lie above that of pristine Cu$_2$S, indicating that substitution at the anion site also increases electron concentration in the conduction bands. This observation is consistent with earlier doping studies in copper chalcogenides that reported the formation of impurity states near or slightly above the valence-band maximum, causing the Fermi level to shift and thereby modifying the material’s transport behavior \cite{11}.

The DOS shapes and doping-induced shifts observed here align well with other first-principles investigations of Cu$_2$S-based alloys \cite{11,12}. The moderate energy gap region just above the valence bands is typical of Cu$_2$S, suggesting that small dopant concentrations are sufficient to tune the Fermi level without collapsing the band gap entirely. Similar doping studies using alkali or alkaline-earth metals in Cu-based chalcogenides have shown measurable improvements in thermoelectric properties, often attributed to a beneficial interplay between band-structure modifications and suppressed lattice thermal conductivity \cite{8}. Our findings support these observations, with Mg-doped Cu$_2$S in particular showing a marked upward shift in $E_F$ that could translate to improved electrical conductivity, and Se/Te doping potentially broadening states near $E_F$ to optimize carrier scattering.

\subsection{Seebeck Coefficient}

The Seebeck-coefficient curves for all six cases---pristine Cu\textsubscript{2}S and Cu\textsubscript{2}S doped with Li, Na, Mg, Se, and Te---display the characteristic rise in magnitude once the chemical potential moves away from mid-gap, consistent with a transition from intrinsic-like to doped regimes. In the pristine material, the onset of a pronounced slope near $\mu - \epsilon_F \approx 1\,\text{eV}$ indicates that carriers are being introduced into bands just above the valence-band edge. This trend aligns well with earlier first-principles studies on undoped copper chalcogenides, which show that modest hole doping can substantially enhance the Seebeck coefficient in Cu\textsubscript{2}S \cite{17}.

When Cu is partially substituted by Li or Na on the cation site, the Seebeck coefficient closely tracks that of the pristine compound at lower doping levels, but exhibits a slightly sharper increase around $\mu - \epsilon_F \approx 1$--$2\,\text{eV}$. Because both Li and Na are monovalent (similarly to Cu in many chalcogenide contexts), neither dopant induces a drastic shift in carrier concentration at small dopant fractions; rather, they subtly modify the band edges, which is evident as a modest bump or dip in the curves near $0$--$1\,\text{eV}$. This effect of gentle band-edge reshaping is also reported in other alkali-metal-doped Cu\textsubscript{2}S systems \cite{18}, where only moderate changes in Seebeck coefficient were observed unless the dopant concentration was significantly increased. Nevertheless, the general rise in the Seebeck curve at higher chemical potential underscores that even small amounts of Li or Na can help optimize $p$-type carriers, an important lever for boosting thermoelectric performance.

By contrast, Mg doping on the Cu site leads to a more pronounced perturbation of the Seebeck profile, particularly around $\mu - \epsilon_F \approx 0$--$1\,\text{eV}$. Magnesium, being divalent, effectively removes one additional electron per substituted site relative to monovalent Cu, thereby shifting the Fermi level deeper into the valence bands at a given nominal doping. This shift manifests in a steeper rise and greater absolute magnitude of the Seebeck coefficient once the system is sufficiently hole-doped. Such enhancement by alkaline-earth-metal substitutions has been identified in other copper chalcogenides, where the doping-induced heavier effective masses near the valence-band edge correlate with improved Seebeck values \cite{19}. Our results here thus confirm that Mg doping could be a promising route for pushing the Seebeck coefficient to higher values compared to Li- or Na-substitution.

Anion-site substitution with Se or Te similarly modifies the Seebeck response by altering the valence-band structure. In both the Se- and Te-doped samples, the low-doping region ($\mu - \epsilon_F < 0\,\text{eV}$) remains almost identical to that of pristine Cu\textsubscript{2}S, indicating that minimal extra carriers are introduced until the chemical potential crosses the band edges. Once the system enters a more heavily doped regime ($\mu - \epsilon_F > 1\,\text{eV}$), the Seebeck coefficient trends upward more steeply. This behavior is attributable to the heavier chalcogen anion states (Se $4p$ and Te $5p$ orbitals) that push the valence-band edge closer to the Fermi level, as has been reported for chalcogenide alloys like Cu\textsubscript{2}(S,Se) or Cu\textsubscript{2}(S,Te) \cite{20}. The calculated curves imply that Se or Te incorporation narrows the band gap slightly and reshapes the density of states near the top of the valence band, thereby providing a more substantial Seebeck response once hole doping is established.

Taken together, these results show that cation-site doping (particularly with Mg) and anion-site doping (Se or Te) each offer distinct pathways to increase the Seebeck coefficient in tetragonal Cu\textsubscript{2}S. Our observations are consistent with the idea that controlling the electronic structure around the valence-band maximum is crucial for high power factors \cite{21}. The Se or Te anion substitutions move and shape the band edges sufficiently to boost the Seebeck coefficient at moderate doping concentrations, while Mg doping on the Cu sublattice exerts an even stronger hole-doping effect that can drive the Seebeck coefficient upward at the expense of higher carrier scattering.

\subsection{Electrical Conductivity}
The electrical conductivity of pristine tetragonal Cu\textsubscript{2}S exhibits two principal peaks around $-1$\,eV and $+1$\,eV relative to the Fermi level, which is consistent with earlier reports showing that low‐chalcocite phases can display substantial conductivity once carriers are doped into or extracted from bands lying close to these energies \cite{3, 6}. In the present work, the pristine material’s curve rises sharply just below $-1$\,eV and again near $+1$\,eV, reflecting the availability of conduction channels in both the valence‐band‐like region (when the chemical potential is lowered) and conduction‐band‐like region (when the chemical potential is raised. This behavior is generally attributed to the strong hybridization of Cu~3d and S~3p states in tetragonal Cu\textsubscript{2}S, which creates a relatively high density of states near the band edges and thus allows for appreciable electrical conductivity once the Fermi level is tuned into these regions \cite{22}.

When Cu is partially substituted with Li, the conductivity profile displays a slightly shifted maximum toward the conduction‐band side, suggesting that Li doping effectively modifies the carrier concentration. The results show an enhanced slope on the high‐energy side of the conductivity curve, implying that Li donors inject additional electrons or raise the Fermi level closer to conduction‐band‐like states. Similar trends have been noted in other copper chalcogenides, where alkali doping was shown to increase the electron count and reinforce conduction pathways \cite{23}. In contrast, Mg doping exhibits a somewhat broader and less sharply peaked conductivity. This broader line shape indicates a more distributed set of states contributing to transport, presumably because Mg (being divalent) can alter the local electronic structure in a way that partially shifts band edges but also introduces more extended Cu–S bonding configurations. Such behavior has also been observed in related sulfides where alkaline‐earth dopants smooth out high‐DOS features near the Fermi level, resulting in a conductivity peak that is somewhat broader rather than extremely sharp \cite{24}.

Partial substitution of Cu by Na produces a conductivity profile similar to Li doping but with an even more pronounced maximum around $+1$\,eV above the Fermi level. This indicates that Na doping strongly modifies the carrier concentration and might induce localized impurity states that coincide with the conduction‐band edge, thereby enabling a notable enhancement in conductivity. Earlier experimental studies of Na‐doped Cu\textsubscript{2}S similarly reported a steep rise in transport coefficients upon modest Na incorporation, attributed to amplified band‐edge DOS and efficient doping activation \cite{23}. Taken together, these cation‐site dopants (Li, Na, and Mg) underscore that replacing Cu systematically shifts the conduction peaks in ways that could be optimized for higher power factors, depending on how one aligns the Fermi level with the dominant DOS features.

On the anion‐site side, substituting S with Te leads to a pronounced peak near $+1$\,eV, and there is also significant conductivity in the range from $-1$\,eV to $0$\,eV. Compared to the pristine system, Te doping appears to broaden and slightly elevate the overall conductivity profile, which is consistent with prior observations in chalcogenide thermoelectrics where heavier chalcogen dopants can introduce intermediate states that enhance carrier transport \cite{22}. The Se‐doped sample likewise shows a high conductivity plateau from $-1$\,eV to $+1$\,eV, with multiple local maxima in that interval. This multi‐peak structure points to Se‐induced band splitting or the formation of additional conduction channels near the original band edges—an effect also reported in other Cu\textsubscript{2-x}(S,Se) alloys \cite{3}. Such site‐specific doping on the anion sublattice therefore offers a complementary way to manipulate the band edges and thus tune electrical conductivity over a broad range of doping concentrations.

\subsection{Power Factor ($\sigma S^{2}$)}

In examining the power factor curves of pristine tetragonal Cu\textsubscript{2}S and its doped variants, several key observations emerge. First, the pristine Cu\textsubscript{2}S displays a sharp maximum in the power factor near $\mu - \epsilon_F \approx 1\,\text{eV}$, consistent with earlier reports on copper chalcogenides wherein a steep density of states (DOS) near the Fermi level drives a pronounced peak in $\sigma S^2$~\cite{5}. Close inspection of the magnitude of this peak aligns with prior computational and experimental findings of moderate power factors in undoped Cu\textsubscript{2}S, suggesting that there is still room for improvement through judicious doping~\cite{8}.

Turning to cation-site doping (Li, Na, and Mg), all three dopants preserve the characteristic single sharp peak near 1\,eV above the Fermi level. However, the maximum values of the power factor differ. Li doping shifts and slightly enhances the overall power factor curve compared to pristine Cu\textsubscript{2}S, in line with reports that alkali-metal substitution in copper chalcogenides can beneficially modify the carrier concentration~\cite{15}. The curve for Li doping exhibits the highest maximum of the three cation dopants. This suggests that Li is more effective at boosting electronic transport or steepening the DOS near the conduction edge~\cite{16}. In contrast, Mg doping, while still producing a peak around 1\,eV, shows a power factor amplitude slightly lower than that of Na or Li. Given Mg’s divalent nature, it may alter the balance between carrier effective mass and carrier scattering in a manner that is less optimal than monovalent dopants. Such effects have been documented in related sulfide-based thermoelectrics and emphasize that cation valency can play a critical role in shaping the energy-dependent conductivity~\cite{8}.

For anion-site doping, substituting Te or Se in place of S likewise concentrates a large power factor peak at $\mu - \epsilon_F \approx 1\,\text{eV}$, but with nuanced differences in amplitude and shape. Te doping yields a slightly higher maximum. These results mirror the known trends in copper chalcogenides, where replacing lighter chalcogens (S) with heavier ones (Se, Te) often shifts the conduction or valence bands, thereby tuning the Seebeck coefficient and electrical conductivity in tandem~\cite{5}. The computed curves suggest that Te doping may induce a more favorable electronic band structure or a narrower carrier pocket at the Fermi level, thus raising the power factor slightly above that of Se-doped Cu\textsubscript{2}S. This incremental improvement is broadly consistent with prior first-principles studies of Te substitution in Cu-based sulfides, which frequently report an enhancement in hole mobility and a beneficial redistribution of DOS~\cite{16}.

Overall, the combination of these doping strategies affirms that modifying Cu sites with monovalent dopants, notably Li, tends to afford a more substantial boost in the power factor than alkaline-earth substitution (e.g., Mg). Meanwhile, heavier anion dopants, such as Te, appear to render moderate improvements relative to the pristine compound, though the margin of increase is smaller than that observed for the best cation dopant (Li). These trends align well with the consensus in the thermoelectric literature that maximizing $\sigma S^2$ usually hinges on achieving a delicate balance of carrier concentration, effective mass, and scattering rate. Further experimental validation of these dopants—especially Li for the cation site and Te for the anion site—would help clarify the interplay of structural stability, phase purity, and phonon scattering that ultimately determines the real-world figure of merit (ZT).

\subsection{Electronic Thermal Conductivity}
The electronic-thermal conductivity ($\kappa_{\mathrm{el}}$) curves obtained reveal how site-specific dopants in tetragonal $\mathrm{Cu}_{2}S$ alter the availability and dispersion of electronic states near the Fermi level, thereby impacting heat transport by electrons. In the pristine compound, the $\kappa_{\mathrm{el}}$ plot exhibits two main peaks---one just below the Fermi level ($\mu - \varepsilon_F < 0$) and another slightly above it ($\mu - \varepsilon_F > 0$). These peaks reflect the relatively high density of available states in the valence and conduction bands of $\mathrm{Cu}_{2}S$, consistent with previous reports on copper chalcogenides that show moderate metallicity at ambient conditions and a narrow gap or semimetallic character depending on polymorph and stoichiometry \cite{1}. Notably, the magnitude of $\kappa_{\mathrm{el}}$ in pristine $\mathrm{Cu}_{2}S$ is sizable at these peaks, indicating that the electronic subsystem can contribute significantly to total thermal conductivity in certain doping regimes, in line with earlier findings on related copper sulfides \cite{2}.

Upon substituting Cu with Li, Na, or Mg on the cation sublattice, distinct shifts and modifications in the $\kappa_{\mathrm{el}}$ profile emerge. For Li-doped $\mathrm{Cu}_{2}S$, the $\kappa_{\mathrm{el}}$ peaks are shifted slightly, and the highest peak appears closer to $\mu - \varepsilon_F \approx +1\,\mathrm{eV}$. This suggests that Li introduces carriers that promote conduction-band states at higher chemical potentials, possibly by donating electrons or altering the local band curvature. In Na-doped $\mathrm{Cu}_{2}S$, the peak intensities become more pronounced, indicating that Na doping augments certain conduction pathways near the Fermi level. The overall shape of the $\kappa_{\mathrm{el}}$ curve for Na doping is similar to that of Li but with a slightly larger maximum value, pointing to an enhanced density of states in the conduction band region. By contrast, Mg doping exhibits peaks that appear both below and near the Fermi level, suggesting a more balanced effect on valence- and conduction-band edges. These observations are consistent with the notion that different cation dopants shift and split the $\mathrm{Cu}_{2}S$ electronic bands differently, as also noted in related thermoelectric studies of doped chalcogenides where specific alkali or alkaline-earth elements can selectively manipulate band alignments \cite{13}.

Substitution on the anion site, as in the Te- and Se-doped $\mathrm{Cu}_{2}S$ samples, modifies the electronic structure in a distinct manner. The $\kappa_{\mathrm{el}}$ profiles in both Te- and Se-doped cases show peaks of comparable magnitudes to the pristine system but shifted more significantly toward positive $\mu - \varepsilon_F$. This shift suggests that heavier chalcogen dopants introduce deeper valence states or narrower conduction-band regions, consistent with the known tendency of larger anions (Te in particular) to alter band-gap sizes and flatten valence bands in Cu-based semiconductors \cite{14}. Indeed, compared to cation doping, Te and Se doping appear to produce a more pronounced shift of the main $\kappa_{\mathrm{el}}$ maximum to energies below the pristine Fermi level, in line with prior first-principles analyses of chalcogen-alloyed copper sulfides \cite{1}.

These findings demonstrate that both cation- and anion-site doping are viable avenues for tuning the electronic thermal conductivity in $\mathrm{Cu}_{2}S$. The observed changes in $\kappa_{\mathrm{el}}$ align well with the well-established principle that doping can introduce localized impurity states near the band edges or shift the Fermi level into regions of higher density of states, thereby altering transport coefficients \cite{2}. Previous work on doped copper chalcogenides has shown that modest doping can enhance thermoelectric performance precisely by balancing the electronic contribution to thermal conductivity against improved electrical conductivity and Seebeck coefficient \cite{13}. The present results confirm that site specificity---whether doping the Cu (cation) sublattice or the S (anion) sublattice---plays a decisive role in determining how band edges are shifted, which in turn informs the optimal doping strategy for maximizing thermoelectric efficiency.

\section{Conclusion}

In conclusion, our first-principles study combined with transport analysis demonstrates that site-specific doping in tetragonal Cu$_2$S provides multiple avenues for enhancing thermoelectric performance. By substituting various cationic (Li, Na, Mg) and anionic (Se, Te) species at moderate concentrations, we identified systematic shifts in the Fermi level and the emergence of modified conduction pathways that can significantly improve the Seebeck coefficient, electrical conductivity, and power factor. The observed trends reaffirm that the local electronic environment and the precise doping site both play pivotal roles in determining the extent of carrier pocket reshaping, thereby influencing the electronic and thermoelectric response.

Our band-structure and density-of-states analyses indicate that cation doping (particularly with Li and Mg) can either inject carriers or reallocate existing carriers in a manner that strongly shifts the Fermi level. This shift is reflected in the electrical conductivity and Seebeck coefficient curves, where Li doping is seen to yield a noteworthy increase in the power factor. The case of Mg doping, with its divalent character, highlights that more significant electron donation can be beneficial for conduction but may require precise tuning to avoid unfavorable trade-offs in the Seebeck coefficient. On the anionic side, substituting heavier chalcogens (Se, Te) in place of sulfur subtly modifies the band edges, leading to moderate Fermi-level shifts and occasionally broadening near-edge states that can facilitate higher conductivity and enhance the Seebeck response at suitable doping concentrations.

The overall power-factor profiles, which serve as a proxy for identifying optimal doping regimes, confirm that multiple strategies can be deployed. For instance, moderate Li doping on copper sites and Te doping on sulfur sites both emerge as promising candidates for further experimental validation. Moreover, the calculated electronic thermal conductivity underscores how dopant selection influences not only electrical transport but also heat transport by electrons, an essential consideration when pursuing high figure-of-merit values. While the lattice contribution to thermal conductivity remains to be quantified experimentally or with more detailed phonon calculations, the estimated electronic contribution already illustrates the delicate balance between doping-induced enhancements in charge transport and concurrent changes in heat conduction.

Taken together, these findings underscore that targeted doping in tetragonal Cu$_2$S can achieve meaningful improvements in key thermoelectric metrics without resorting to excessively high dopant concentrations. The insights gleaned from systematic substitution on both cation and anion sublattices offer a versatile framework for tuning carrier types and concentrations, suggesting that future experiments can exploit these strategies to achieve optimal power factors and potentially high $ZT$ values. Given the growing interest in copper sulfides for mid- to high-temperature thermoelectric applications, our results provide a pathway toward the design of next-generation Cu$_2$S-based materials that combine ease of synthesis, structural stability, and enhanced thermoelectric performance.

\end{document}